\documentclass[aip,rsi,reprint, numerical]{revtex4-1}

\usepackage{graphicx}

\begin{document}


\title{Modular cryostat for ion trapping with surface-electrode ion traps} 



\author{Grahame Vittorini}
\email[]{grahame.vittorini@gatech.edu}
\affiliation{School of Physics, Georgia Institute of Technology, Atlanta, GA 30332 USA}

\author{Kenneth Wright}
\altaffiliation{current address: Joint Quantum Institute, University of Maryland, College Park, MD  20742 USA}
\affiliation{Georgia Tech Research Institute, Atlanta, GA 30332 USA}

\author{Kenneth R. Brown}
\affiliation{School of Physics, Georgia Institute of Technology, Atlanta, GA 30332 USA}
\affiliation{Schools of Chemistry and Biochemistry, and Computational Science and Engineering, Georgia Institute of Technology, Atlanta, GA 30332 USA}

\author{Alexa W. Harter}
\affiliation{Georgia Tech Research Institute, Atlanta, GA 30332 USA}

\author{S. Charles Doret}
\affiliation{Georgia Tech Research Institute, Atlanta, GA 30332 USA}

\date{\today}

\begin{abstract}
We present a simple cryostat purpose built for use with surface-electrode ion traps, designed around an affordable, large cooling power commercial pulse tube refrigerator.  A modular vacuum enclosure with a single vacuum space facilitates interior access, and enables rapid turnaround and flexibility for future modifications.  Long rectangular windows provide nearly 360 degrees of optical access in the plane of the ion trap, while a circular bottom window near the trap enables NA 0.4 light collection without the need for in-vacuum optics.  We evaluate the system's mechanical and thermal characteristics, and we quantify ion trapping performance by trapping $^{40}$Ca$^{+}$, finding small stray electric fields, long ion lifetimes, and low ion heating rates.
\end{abstract}


\maketitle 

\section{Introduction}
Trapped atomic and molecular ions represent a promising system for quantum information processing,\cite{Wineland11, Duan10} while also offering opportunities for metrology\cite{Rosenband08} and precision measurement.\cite{Thompson90, Schiller05, Leanhardt11, Beloy11, Kozlov11}  Recent advances have led to smaller trap footprints as researchers pursue trap designs and fabrication technologies that can hold larger numbers of ions.  Particularly noteworthy in this regard is the introduction of the surface-electrode ion trap.\cite{Chiaverini05, Seidelin06} This geometry is well suited to miniaturization and microfabrication as all of the trap electrodes are located in a single plane.  Unfortunately, planarizing the trap comes at the cost of reduced trapping pseudopotential depth, making surface-electrode traps susceptible to ion loss or reordering of ion chains due to collisions with background gas.  Furthermore, small ion-electrode distances in miniaturized traps contribute to rapid anomalous heating of the trapped ions,\cite{Turchette00, Deslauriers06} reducing gate fidelites and limiting coherence times.  

Placing a surface-electrode trap in a cryogenic environment can significantly mitigate these drawbacks.  Ion lifetimes are increased due to improved vacuum and reduced collision energies, while Johnson noise and anomalous heating are both suppressed.\cite{Deslauriers06, Labaziewicz08}  Furthermore, reduced outgassing at cryogenic temperatures permits the use of a wide range of materials not appropriate for use in room temperature ultra-high vacuum (UHV) chambers.  Cryogenic environments may also facilitate coupling of trapped ion systems to other quantum architectures, for example via superconducting striplines.\cite{Tian04}

Despite these advantages, only a handful of low temperature radio frequency (RF) Paul trapping systems have been built, in part because antithetical design considerations make most off-the-shelf cryostats poorly suited to ion trapping.  Liquid helium based systems\cite{Poitzsch96, Okada01, Antohi09, Brown11} are expensive and impractical to operate due to the large heat loads contributed by oven-based ion sources,  the multitude of electrical connections required for trap control, and the radio frequency trap drive.  Closed-cycle systems\cite{Antohi09, Sage12, Gandolfi12, Schwarz12} eliminate cryogen use but have required expensive helium gas heat exchangers to reduce vibration to a level appropriate for high-fidelity gate operations and high-resolution imaging.  Finally, laser cooling and trapped ion fluorescence collection require optical access from many directions and over large solid angles, complicating radiation shielding of the interior of any cryostat.  

In this work we present a compact cryostat that addresses these conflicting requirements.  The cryostat utilizes a standard commercial cryocooler and a simple vibration isolation assembly, providing increased cooling power at reduced cost relative to helium heat exchangers.  The vacuum enclosure and 50~K radiation shield are highly modular, facilitating rapid turnaround ($<$24 hours from breaking vacuum to retrapping ions) while adding flexibility for future modifications to incorporate new trap designs or to change laser access.  We demonstrate that excellent vacuum can be achieved without any 4~K enclosure by differentially pumping the interior of the 50~K shield with charcoal getters.  By eliminating this extra layer of shielding we improve laser access in the trap plane and enable fluorescence collection over a large solid angle (NA 0.4) without requring in-vacuum optics.   

\section{Experimental Apparatus}
\begin{figure}
\includegraphics[width=8.0 cm]{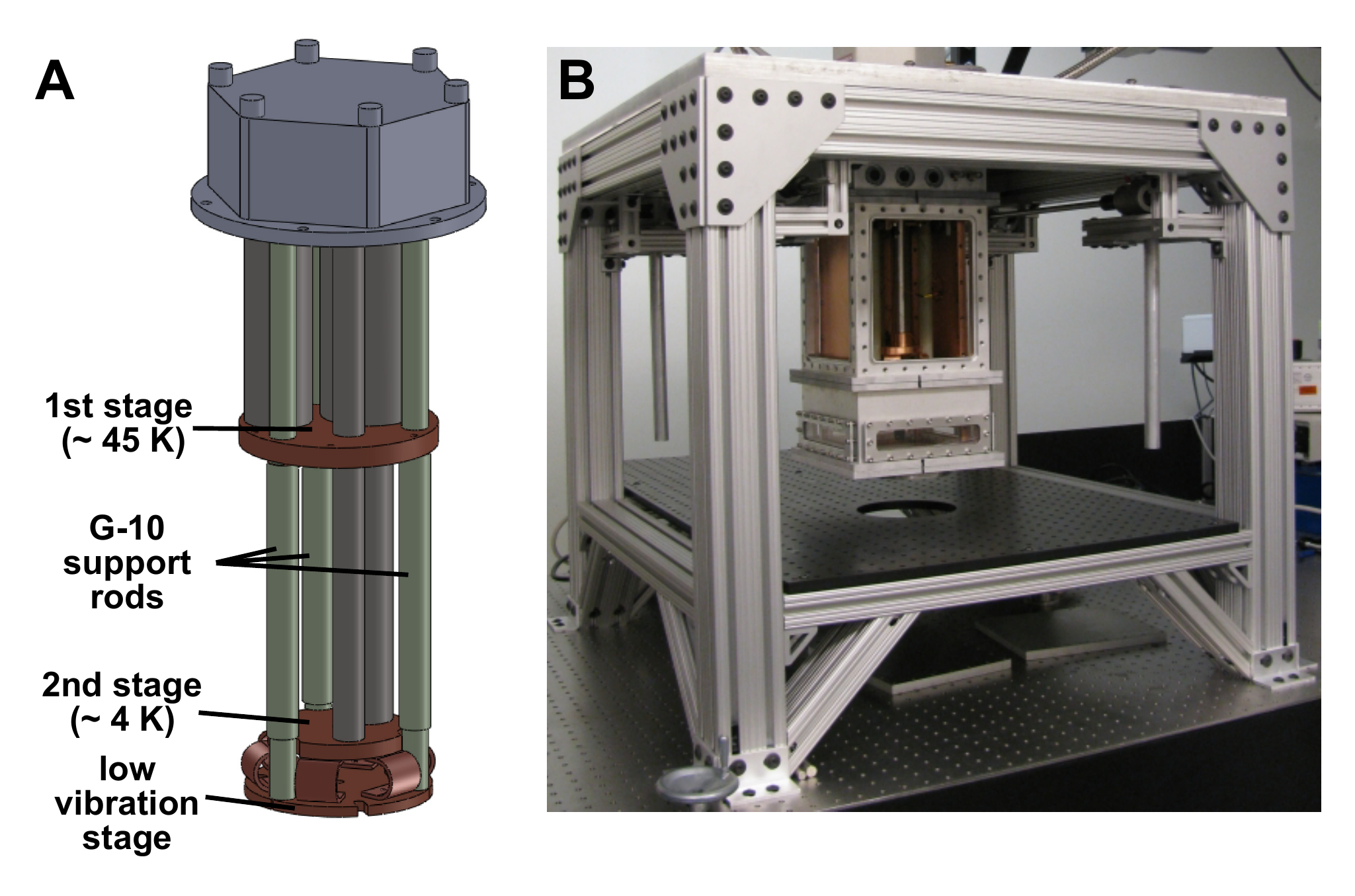}
\caption{(a) The Cryomech PT407 cryocooler, with low vibration stage (remote motor and helium reservoirs not shown).  Ion traps are anchored to the low-vibration stage, which is thermally linked to the 4~K second stage of the cryocooler.  (b) The cryostat (with side panels removed) and superstructure.}
\label{fig:coldhead}
\end{figure}

The core of the experimental apparatus is a pulse-tube cryocooler (Cryomech PT407; Fig.~\ref{fig:coldhead}a). This two-stage cryocooler is specified to provide 0.6~(22)~W of cooling power at 4.2~(55)~K and achieves a base temperature of 2.8 K with no applied heat load.  We use the reduced vibration model of the PT407, for which the cold-head motor and helium reservoirs are mechanically isolated by flexible gas lines.  
The cold-head is attached to an aluminum mounting plate which sits on a rigid superstructure made of extruded aluminium framing pieces (Fig.~\ref{fig:coldhead}b).  During operation the mounting plate is rigidly attached to the superstructure; however, four 100~kg capacity screwjacks (Joyce-Dayton WJ250) allow the mounting plate and cold-head to be raised by up to 30 cm to provide access to the sample stage during trap installation.

\subsection{Vacuum Enclosure} 
Nearly all surfaces become effective cryopumps when cooled to cryogenic temperatures.  Cryopumping may be enhanced by incorporating large surface area adsorbents (getters of activated charcoal, zeolites, or metal sinters), leading to pumping speeds orders of magnitude larger than those achieveable in room temperature vacuum systems.  As outgassing is also negligible at low temperatures, differential pumping is adequate to achieve exceptional vacuum inside a cryostat even while relaxing the requirements on the vacuum enclosure.  

\begin{figure}
\includegraphics[width=8.0 cm]{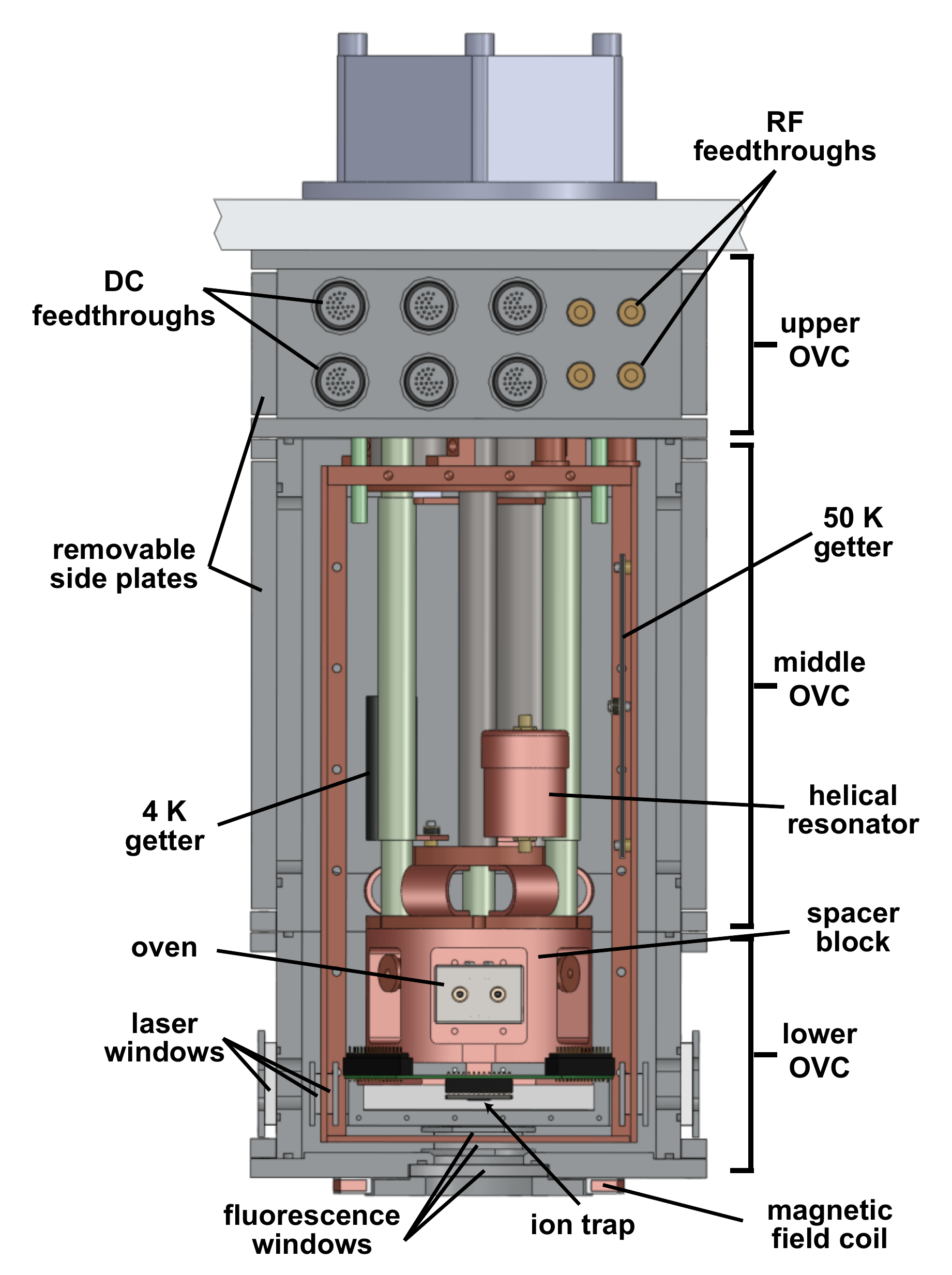}
\caption{Section view of the cryostat including the cold-head, vacuum enclosure, radiation shield, and trap.}
\label{fig:OVC}
\end{figure}

With this in mind, our outer vacuum can (OVC) is designed to emphasize modularity, ease-of-assembly, and optical access.  In lieu of the stainless steel ConFlat (CF) components used in some other systems,\cite{Antohi09, Sage12, Schwarz12} the OVC is aluminum with viton O-ring seals.  This simplifies machining, reduces weight, and eliminates magnetic materials.  The enclosure is composed of three sections (Fig.~\ref{fig:OVC}), each made from a length of extruded 20~$\times$~20~cm square tubing with a 1.25~cm wall thickness (Metals Depot T38812).  This square extrusion allows for more tightly-packed feedthroughs and larger windows than would be possible with a cylindrical geometry.
To simplify construction, vacuum seals are formed via O-ring grooves cut directly into the end face of each section of extrusion, eliminating the need for any vacuum-tight welds.  Flanges welded to the outside of each section at the top and bottom accomodate bolts for assembly.    

The top section of the enclosure forms a short collar containing electrical and vacuum feedthroughs, while the bottom and middle sections may be removed to provide access to the ion trap and the region between the cryocooler stages, respectively.  Interfaces between the sections are approximately matched to the heights of the first and second stages of the cryocooler, allowing piece-wise disassembly to access critical components without ever breaking electrical connections routed from the upper feedthrough collar.  The design also incorporates removable side panels on all blank faces of the OVC.  These panels afford rapid entry to the vacuum space and also increase modularity; a single panel may be changed or replaced (for example, to add additional electrical feedthroughs) without modifying other parts.  Three such panels are mounted on the upper OVC, while the fourth (fixed) face contains electrical feedthroughs.  Four large panels on the middle OVC open to the remainder of the vacuum space.

\subsection{Radiation Shield and Differential Pumping}
Within the vacuum enclosure a radiation shield made of 3~mm thick copper (OFHC, alloy 10100) is suspended from the aluminum mounting plate by four threaded G-10 rods.  This layer is thermally anchored to the first stage of the cryocooler, reaching a temperature of 50~K when cold.  As with the vacuum enclosure, the heat shield is modular, consisting of side panels mounted to corner posts.  Sub-panels on the front and back of the radiation shield are matched to the side ports of the middle OVC to further simplify interior access.  

In addition to reducing the blackbody heat load, the radiation shield impedes gas flow between the room temperature vacuum enclosure and the cold interior, allowing the interior to be differentially cryopumped to ultra-high vacuum.  To increase pumping speed we utilize two getters inside the shield - one anchored to the shield itself, and a second on the 4~K stage of the cryocooler.  Both are made from activated charcoal (Calgon Carbon) epoxied to copper plates via Stycast 2850FT epoxy.  Nichrome resistive heaters attached to each sorb may be used to speed the desorption of water during initial room temperature pumpout or to regenerate the getters when the system is cold. 

\subsection{Vibration Isolation}
Despite the fact that all of the cryocooler's moving parts are mechanically divorced from the cold-head, residual vibration of the low temperature components remains due to the periodic motion of pressurized helium in the thin-walled stainless steel pulse tubes, as well as their differential thermal contraction.  To reduce the motion of the ion trap, the trap is mechanically isolated from the 4~K stage of the cryocooler by mounting it to an auxiliary sample stage (Fig.~\ref{fig:coldhead}).  This stage is formed from a copper mounting plate that is mechanically anchored to the stainless steel room temperature flange of the cold-head via three G-10 rods, thus rigidly attaching it to the stiff aluminum mounting plate and superstructure.  The G10 rods are thermally anchored to the cryocooler's first stage by flexible copper braids, while six stacked copper foil bridges anchor the copper plate to the 4~K stage.  A massive ($\sim$~5~kg) cylindrical copper block (OFHC, alloy 10100) is anchored to the copper mounting plate.  This block may be modified or replaced to incorporate different trap designs; its large size increases the thermal mass of the system to dampen temperature fluctuations while also reducing the amplitude of any remaining vibration.   

\subsection{Optical Access}
\begin{figure}
\includegraphics[width=8.0 cm]{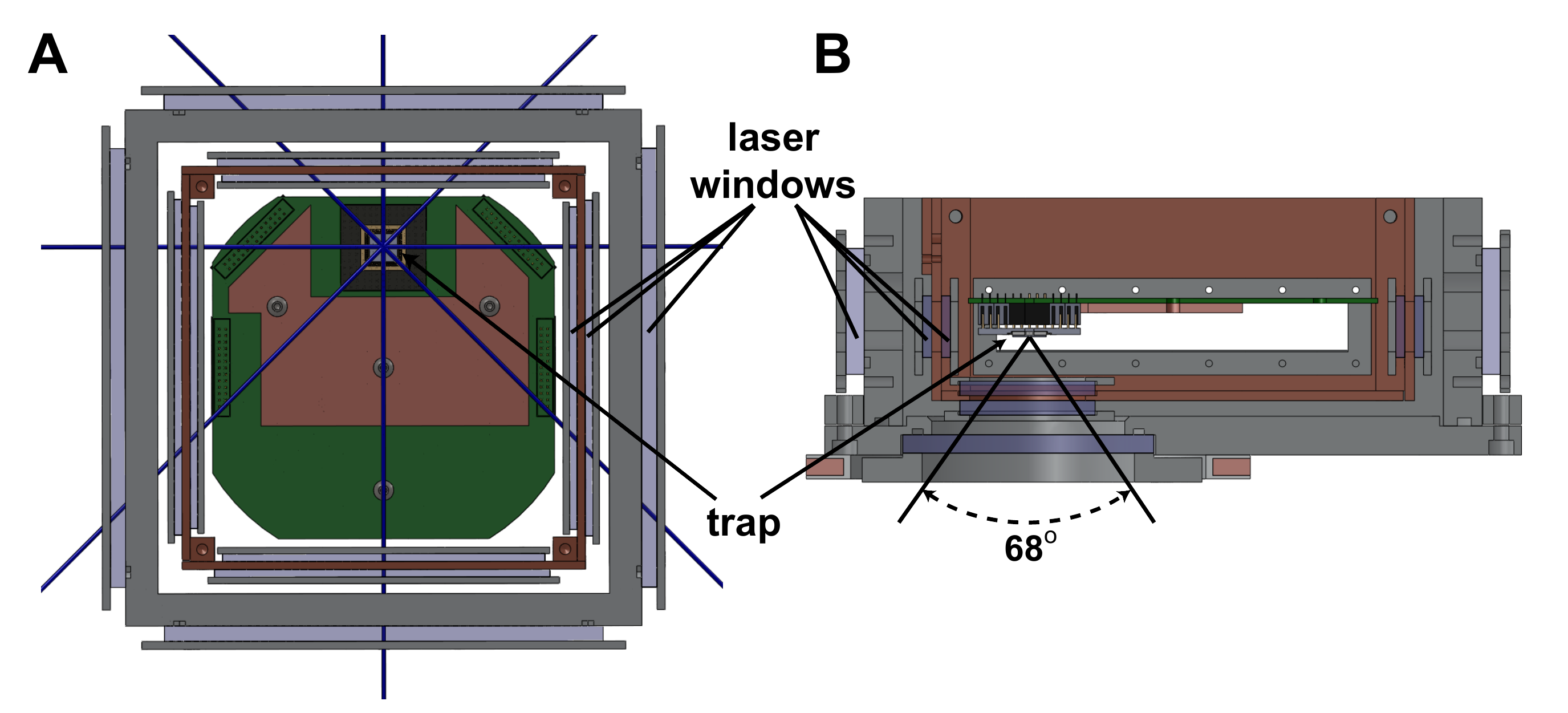}
\caption{Cross-sections showing optical access to the trap.  (a) Side view.  (b) Bottom view, with typical laser paths indicated.}
\label{fig:optical_access}
\end{figure}

Five windows located on the lower OVC provide optical access to the ion trap.  Large rectangular windows (Esco Products) on the side faces of the OVC allow lasers to be aligned across the trap's surface.  Such rectangular windows permit laser entry from almost any direction in the trap plane, offering greater flexibility for laser alignment than commodity components such as CF flanged spherical octagons.  A set of windows at 50~K matched to those on the OVC let light through the radiation shield.    We use two windows on each panel of the shield, sandwiched against the inside and outside faces (Fig.~\ref{fig:OVC}), allowing the outer window to thermally shield the inner one.  The large aspect ratio of the windows is particularly advantageous, providing wide-angle laser access while maintaining a small distance between the windows' centers and their heat-sunk perimeter.  To ensure access along key diagonal beam paths (which would otherwise be blocked by the corners of the square OVC), the trap is shifted off-center in the cryostat (Fig.~\ref{fig:optical_access}a).
All side windows are W anti-reflection coated at 400 and 800~nm, reducing reflection to $<1\%$ per face at all laser wavelengths.  An optical breadboard mounted to the extruded aluminum superstructure holds optics for aligning lasers across the trap.

Ion fluorescence is collected through an additional stack of circular windows located directly underneath the ion trap, V-coated at 397~nm for detection of Ca$^{+}$ fluorescence.  The 300~K window (6.25~cm clear aperture) is recessed into the OVC  bottom plate to mimimize the distance between room temperature light collection optics and the ion trap, allowing for light collection over a large solid angle of nearly 2~sr (Fig.~\ref{fig:optical_access}b).  As such, efficient fluorescence collection is feasible without the use of in-vacuum optics.  We use a home-built NA~0.4 lens, collecting $\sim$~4$\%$ of the ions' fluorescence.  The lens is attached to the OVC bottom via a 2-axis translation stage (OWIS KT150), so that the optical axis of the imaging system can be carefully aligned to the trapped ions to eliminate vignetting and minimize coma. The focus is separately controlled by adjusting the camera position along the optical axis.

\subsection{Electrical Components and Connections}
\begin{figure}
\includegraphics[width=8.0 cm]{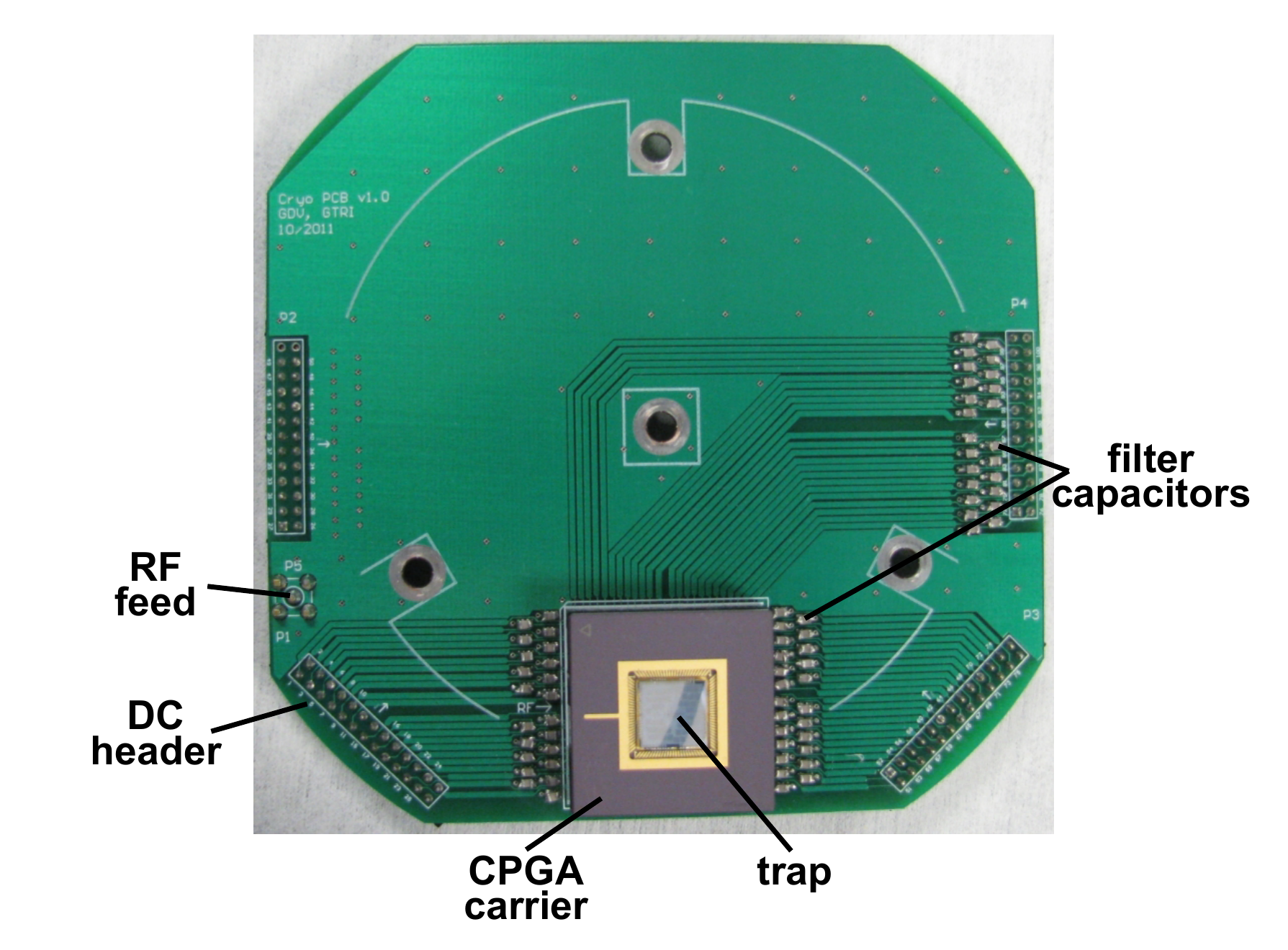}
\caption{The trap socket (with trap installed) and fan-out PCB used to couple DC and RF voltages to the trap electrodes.}
\label{fig:PCB}
\end{figure}

The cryostat is outfitted to accept traps affixed to commercial 100-pin ceramic-pin-grid-array (CPGA) chip carriers, standardizing the installation of traps with varying electrode layouts.  Carriers are mounted in a 100-pin socket soldered to a fan-out PCB (Fig.~\ref{fig:PCB}) made of FR-4 laminate, with each pin filtered by a 33~nF capacitor to ground.  The capacitors are made of NP0 ceramic and show a negligible temperature dependence during cooldown to 4~K.
One pin in each quadrant is grounded at the socket, while the remaining twenty-four pins are routed to a shrouded rectangular header on the perimeter of the PCB.  These headers are fed via wiring from 24-pin O-ring sealed electrical feedthroughs (Fischer DBPE 105 A093-89) in the upper OVC.  36 AWG quad-twist phosphor-bronze wire (Lakeshore WQT-36), heat sunk to the cryocooler's first stage, carries DC voltages to a secondary heat sink at 4~K.  Here the wiring transitions to 32 AWG copper, allowing the DC connections to help heat sink the CPGA and individual trap electrodes.  Two additional 24-pin feedthroughs provide wiring for thermometers (Lakeshore DT-670B1-CU), heaters, the neutral atom oven source, and any future electrical accessories.

The upper OVC also features four SMA feedthroughs (Pasternack PE9184) for introducing RF signals.  Low thermal conductivity stainless steel coaxial cables (Lakeshore CC-SS) carry signals to panel-mount bulkhead SMA connectors attached to the radiation shield, heat sinking the cables at 50~K.  Signals continue from this point via stranded copper coaxial cable (Lakeshore CC-SC).  Two lines connect to the trap's RF electrodes via a compact inductively-coupled helical resonator.\cite{Siverns12}  One line feeds the RF drive voltage, while the second is an optional DC bias voltage (usually grounded).  The resonator is constructed from a 2.65" length of 2" copper pipe (alloy 12200), designed for a resonant frequency of approximately 50~MHz when loaded by the PCB, socket, and trap, with quality factors of 80~(315) at 300~(4)~K, respectively.  One of the remaining two RF lines is connected to an antenna used to monitor applied RF voltages, and the second to a copper ground screen mounted over the trap that shields stray electric fields and can be driven for secular mode frequency measurements.

\subsection{Oven}
Ion traps are commonly loaded by photoionizing atoms from a flux of neutrals sourced by a resistively heated oven.  Depending on the species this can require temperatures well in excess of 1000~K, thus ovens impose a significant thermal load (typically several watts) when energized.  Other methods, such as laser ablation\cite{Antohi09} or transfer from a magneto-optical trap (MOT),\cite{Sage12} offer alternatives with lower heat loads more appropriate for a cryogenic environment.  However, laser ablation requires an additional pulsed laser and lacks the species-selectivity of photoionization unless used carefully in a regime of low pulse energy,\cite{Hendricks07} and for many experiments the efficient trap loading offered by a MOT does not merit the increase in experimental complexity.  Fortunately, suitable thermal shielding and heat sinking can mitigate the degree to which an oven source heats the ion trap, making ovens a viable option in cryocooler-based systems for which there is no concern over liquid helium boiloff.

\begin{figure}
\includegraphics[width=8.0 cm]{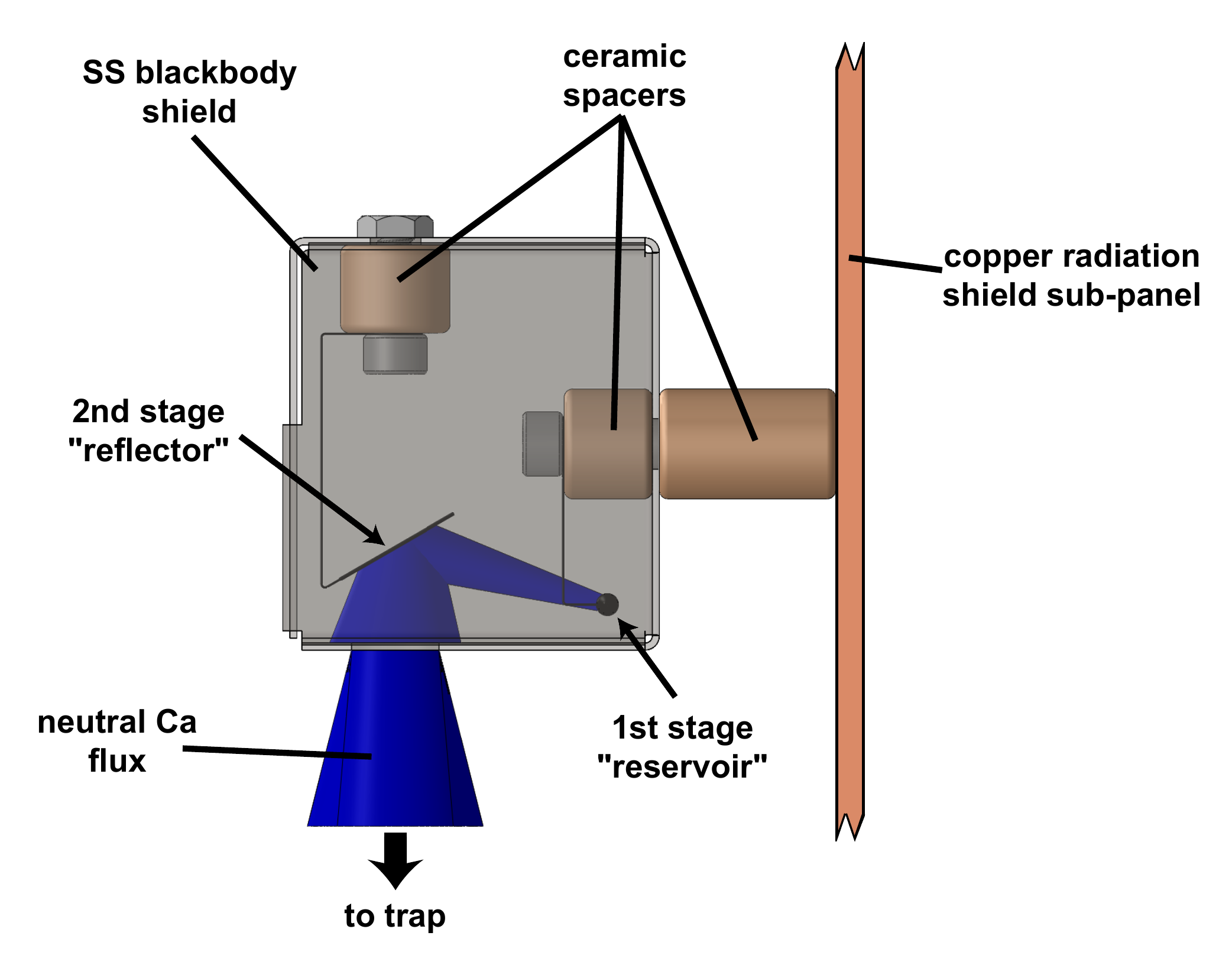}
\caption{The two-stage oven.  Applying current to the reservoir and reflector projects calcium flux onto the trap without risking oxide dust or fragments falling to the trap below.}
\label{fig:oven}
\end{figure}

Our oven is designed for use with backside-loaded ion traps, and thus is situated behind the trap mounting socket.  This puts the oven above the trap, so we use a two-stage design\cite{oven} (Fig.~\ref{fig:oven}) to minimize the chance of oxide and dust falling from the oven onto the trap below.  In this design, a first stage reservoir of calcium metal is resistively heated, producing a flux directed \emph{away} from the trap.  A second stage reflector, also resistively heated, redirects this flux towards the trap loading slot.  The source is made from stainless steel (SS) capillary tube (125~$\mu$m wall thickness, 3~mm OD), filled with calcium metal and crimped/spot-welded on each end to short lengths of 125~$\mu$m diameter SS wire.  A small slit cut circumferentially allows neutral flux to escape towards the reflector -- a small square of 50~$\mu$m thick SS foil spot welded to SS leads.  Both oven stages are anchored to ceramic posts inside a SS sheet metal box which shields emitted blackbody radiation and helps with differential pumping.  This box is mounted to a subpanel on the copper radiation shield, dissipating the heat load to the cryocooler's first stage while minimally heating the ion trap.  This panel may be easily removed and the oven stored elsewhere to minimize oxidation of the calcium metal when the cryostat is opened.  In use the two stages are driven with 2~A and 1~A, respectively, dissipating a total  power of 4~W.

\section{System Characterization}

\subsection{Trap specifics}
\begin{figure}
\includegraphics[width=8.0 cm]{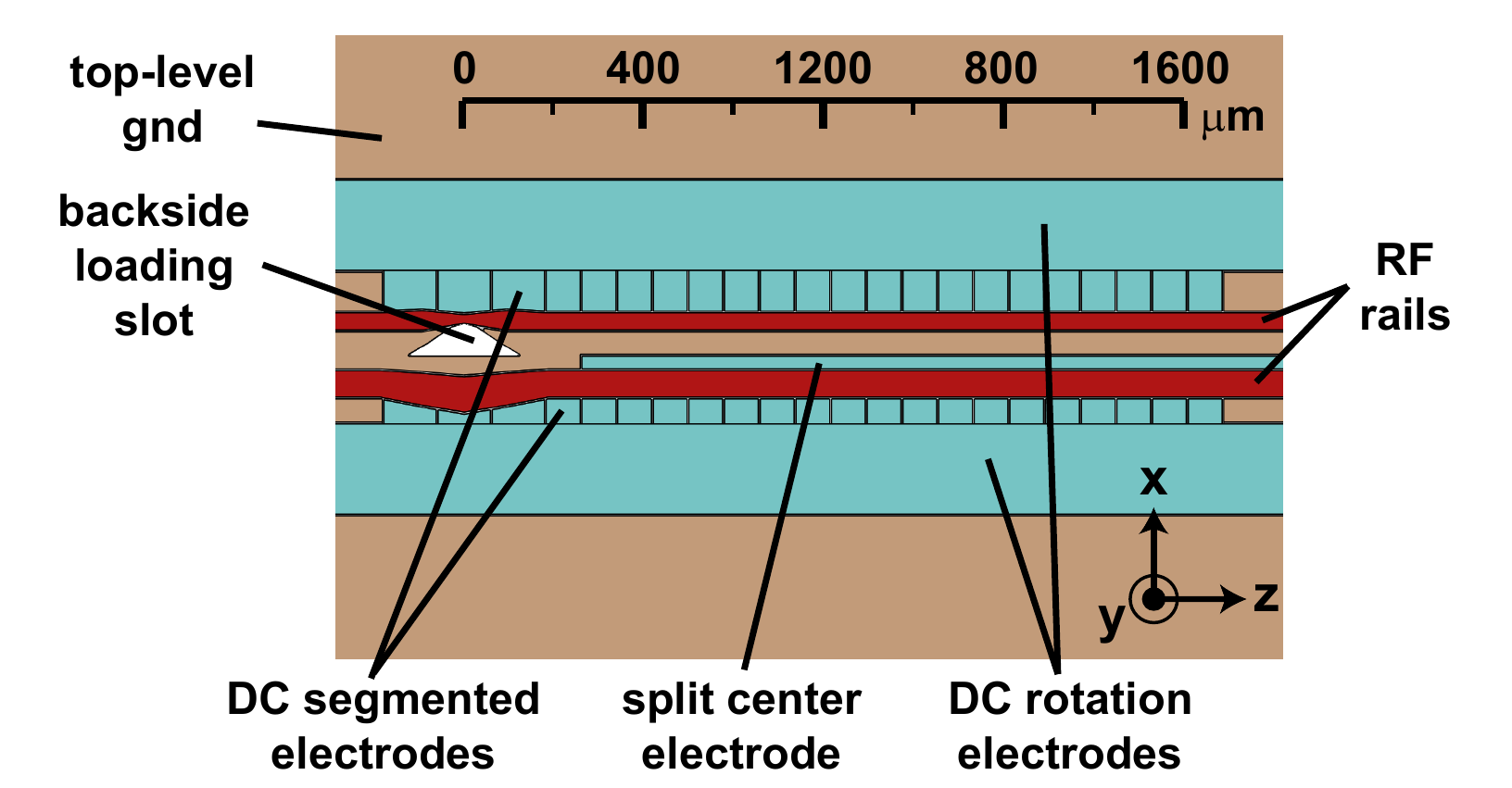}
\caption{The GTRI ``Gen II" surface electrode trap (figure from reference [26]).}
\label{fig:trap}
\end{figure}

To date we have trapped ions in four different surface-electrode traps.  Data presented here were collected using a GTRI ``Gen II" trap\cite{SMITreport, Doret12} (Fig.~\ref{fig:trap}), originally fabricated under Phase II of the Scalable Microfabricated Ion Trap (SMIT II) program.  This aluminum-on-SiO$_2$ trap features a small loading slot, 60~$\mu$m ion height, and forty-four DC electrodes that allow for the fine tuning of control potentials.  The trap is epoxied (EPO-TEK H21D) to an alumina spacer mounted on a CPGA carrier.  Electrical connections to each electrode are made by two 25~$\mu$m diameter 99\% Al / 1\% Si wirebonds; these wirebonds are also the primary thermal anchoring for the trap.

\subsection{Cooldown and Trap Temperature}

To cool the cryostat, we begin by pumping on the system with a turbopump while gently heating the charcoal getters ($\sim 320$~K) to empty them of adsorbed water vapor.  Pumping overnight brings the chamber pressure to approximately $5 \times 10^{-5}$~mbar.  Energizing the cryocooler then rapidly cools the interior of the cryostat; the ion trap temperature drops below 10~K within five hours, with temperatures fully equilibrating ($T_{\text{trap}}=5.5$~K) in approximately twelve hours.  Once cold, cryopumping reduces the chamber pressure outside of the copper radiation shield to approximately $5 \times 10^{-7}$~mbar, with lower pressures inside the differentially pumped volume within.

\begin{figure}
\includegraphics[width=8.0 cm]{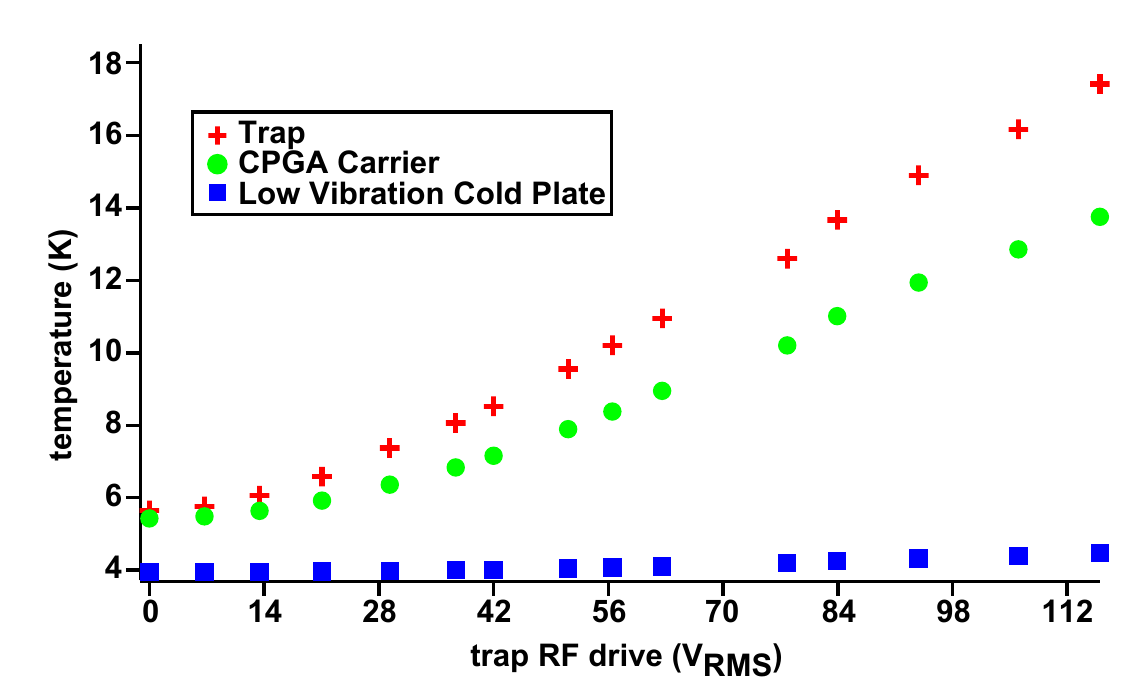}
\caption{Thermal performance as a function of RF drive voltage.  Trap temperatures are measured on the top ground plane at the periphery of the trap chip; thermal modeling suggests that the RF electrodes are 10-20~K higher in temperature.}
\label{fig:trap_temps}
\end{figure}

During normal operation, the RF electrodes on the trap are driven at $\sim 50$~MHz to create a trapping pseudopotential.  Based on the increase in temperature of the cryocooler's second stage, we estimate that a typical RF drive of 70~$V_{\text{RMS}}$ dissipates approximately 100~mW among the resonator, PCB, CPGA carrier, and trap.  Thermometry mounted on the top ground plane of a diagnostic trap indicated that application of this RF drive warmed the trap by approximately 6~K (Fig.~\ref{fig:trap_temps}), dominated by a thermal gradient between the CPGA carrier and the low-vibration cold-plate.\cite{heatsink}  Energizing the oven warms the trap by an additional $\sim$~1~K due primarily to blackbody radiation. No thermometry is mounted on the trap during normal operation. However, temperatures are monitored at the PCB, spacer block, 4~K and 50~K getters, and the cryocooler's first and second stages.

\subsection{Vacuum and Trapping Lifetime}

Despite the poor vacuum in the room temperature region outside of the copper radiation shield, the vacuum around the ion trap is excellent.  The 50~K wall temperature of the radiation shield suppresses line-of-sight outgassing toward the ion trap, so the gas load is dominated by flow from the 300~K exterior of the cryostat through small gaps in the shield.  Fortunately, all gases with non-negligible vapor pressures at 50~K have binding energies on charcoal exceeding 100~K,\cite{Vidali91} and are effectively adsorption pumped onto the cryocooler's second stage and the 4~K charcoal getter.  

\begin{figure}
\includegraphics[width=8.0 cm]{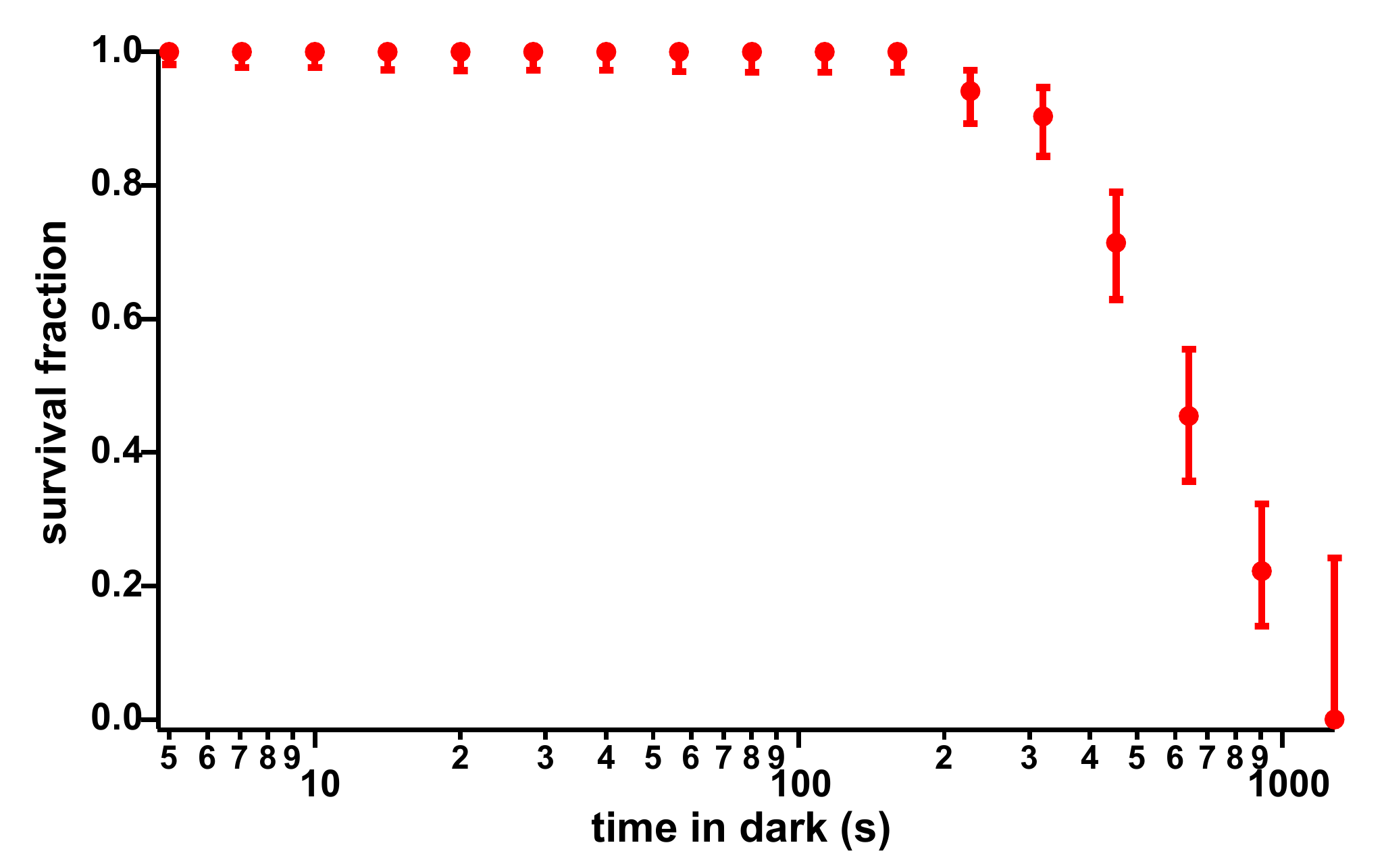}
\caption{Single ion lifetime in the absence of Doppler cooling in a harmonic ($\omega_{z} = 1$~MHz) well with calculated trap depth 32 meV.}
\label{fig:ion_lifetime}
\end{figure}

To measure the quality of the vacuum, we trap single $^{40}$Ca$^{+}$ ions and measure their lifetime in the trap.  Calcium trapping is reviewed elsewhere,\cite{Haffner08} with particulars of our optics and electronics similar to Ref. [26].  Briefly, neutral calcium flux from the resistively heated oven is photoionized, and ions are then laser cooled on the $^{2}$S$_{1/2} \rightarrow ^{2}$P$_{1/2}$ transition at 397~nm; a second laser at 866~nm repumps the ion from a metastable $^{2}$D$_{3/2}$ state.  Doppler-cooled ions survive in the trap for many hours, so we instead measure the lifetime of the ion in the absence of cooling light.  After cooling the ion, we block the 397~nm laser for successively longer intervals, checking for ion survival after each interval with a fluorescence measurement.  The trap is reloaded each time an ion is lost, and after a pause to ensure that the oven has cooled and the system returned to its steady-state conditions, the experiment is repeated many times to obtain ample statistics.  We find a lifetime of approximately 550~s (Fig.~\ref{fig:ion_lifetime}) for a well-depth of 32~meV, significantly longer than the 45~second lifetime measured in a room temperature experiment at $10^{-11}$~mbar.\cite{Doret12} 

\subsection{Vibration}

\begin{figure}
\includegraphics[width=8.0 cm]{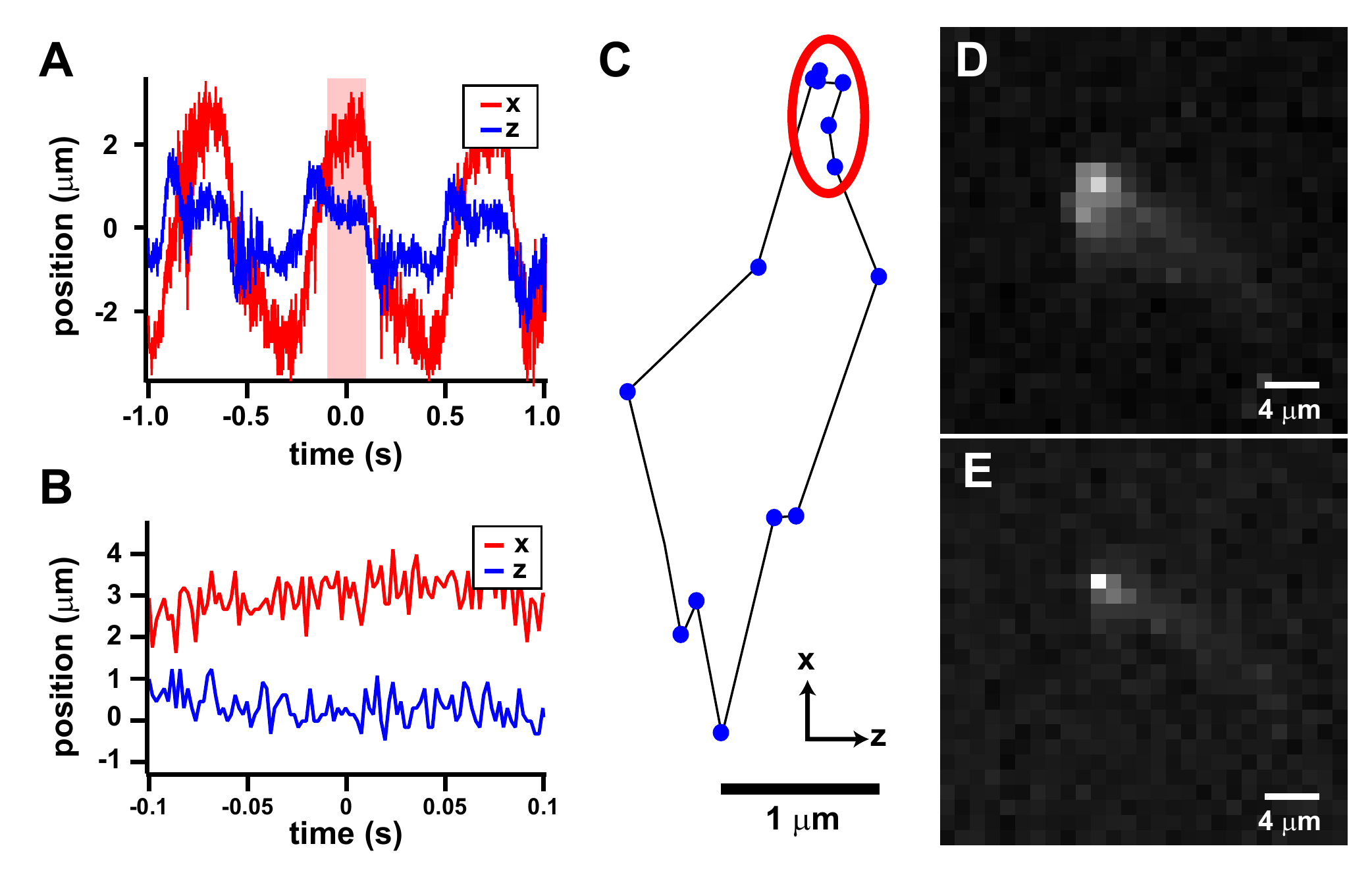}
\caption{Motion of the low-vibration stage.  (a, b) Cryostat vibration in the plane of the ion trap as measured using a razor blade, with directions matching the trap axes indicated in Fig.~\ref{fig:trap}.  (b) shows the shaded region from (a).  High frequency noise is due to detection electronics, and motion is negligible along $\hat{z}$.  (c) Ion motion mapped by triggered CCD fluorescence images.  Each point is separated in time by 50 ms.  The circled cluster of points approximately matches the shaded region in (a).  (d, e) Fluorescence images of the ion averaged over one cycle of the cryocooler vibration or triggered during the low-motion window, respectively.  The ``comet tail" extending down and to the right is due to optical aberration rather than vibrations.}
\label{fig:motion}
\end{figure}

One major drawback to closed-cycle cryocoolers is the vibration induced by the motion of helium gas through the cold-head during the pulse cycle.  For ion trapping this motion leads to blurring of ion fluorescence images as well as phase noise on qubit rotations.  In our system these vibrations are reduced due to the mechanical decoupling of the low-vibration stage.  However, some vibration remains due to minute deformations of the cold-head's SS top flange which are transmitted down the G-10 support rods of the low-vibration assembly to the isolated cold-plate.\cite{vibration}  We quantify this residual vibration by focusing a laser beam past a razor blade mounted on the low vibration stage and measuring the time variation of the transmitted laser power.  With an appropriate focus and a suitable mapping of the detector signal to beam position, sub-micron measurement of the blade position is possible (Fig.~\ref{fig:motion}a,b). 

We also measure the position of the ion directly as a function of the motional cycle of the cryocooler by imaging ion fluorescence onto a CCD camera.  Using a full-bridge strain gauge sensor we generate a TTL trigger signal with which we can reference illumination of the ion to the motional cycle.  The sensor is mounted to a cantilever which bends due to motion of the flex line between the cold-head and remote motor.  We map out the ion motion by fitting a Gaussian to the CCD image to extract the ion's position (Fig.~\ref{fig:motion}c).  These data agree with the razor blade measurements; both methods indicate a total motion of approximately 4.5~$\mu$m peak-to-peak parallel to the trap surface. Measurements with the razor blade indicate motion of $<$1~$\mu$m peak-to-peak perpendicular to the trap plane. However, by triggering experimental data acquisition to coincide with a low-motion window of the cryostat's motional cycle, we can greatly reduce the effective vibration to $<$1~$\mu$m~(Fig.~\ref{fig:motion}d,e). This value represents a significant improvement over the 12~$\mu$m peak-to-peak motion of the PT407 without vibration isolation\cite{vibration} or the $\sim$15~$\mu$m peak-to-peak motion of a large, unstabilized bath cryostat.\cite{BrownPC} However, it is significantly larger than the $<$104~nm motion reported for a closed cycle system using a helium-gas heat exchanger.\cite{Antohi09} Further vibration reduction is possible if the isolated cold-plate were anchored to the stiff aluminum mounting plate instead of the room temperature flange of the cryostat.

\subsection{Measurement and Compensation of Stray Electric Fields}

Despite careful surface-electrode trap fabrication, trapped ions often experience forces due to stray electric fields originating from charging and contamination of the trap and surrounding environment.  These fields can cause excess micromotion, reducing Doppler cooling efficiency and possibly increasing ion heating rates.\cite{Leibfried03}  It is thus important that these fields be characterized and controlled; following measurement, stray fields can be cancelled through application of suitable potentials to the trap electrodes.  

\begin{figure}
\includegraphics[width=8.0 cm]{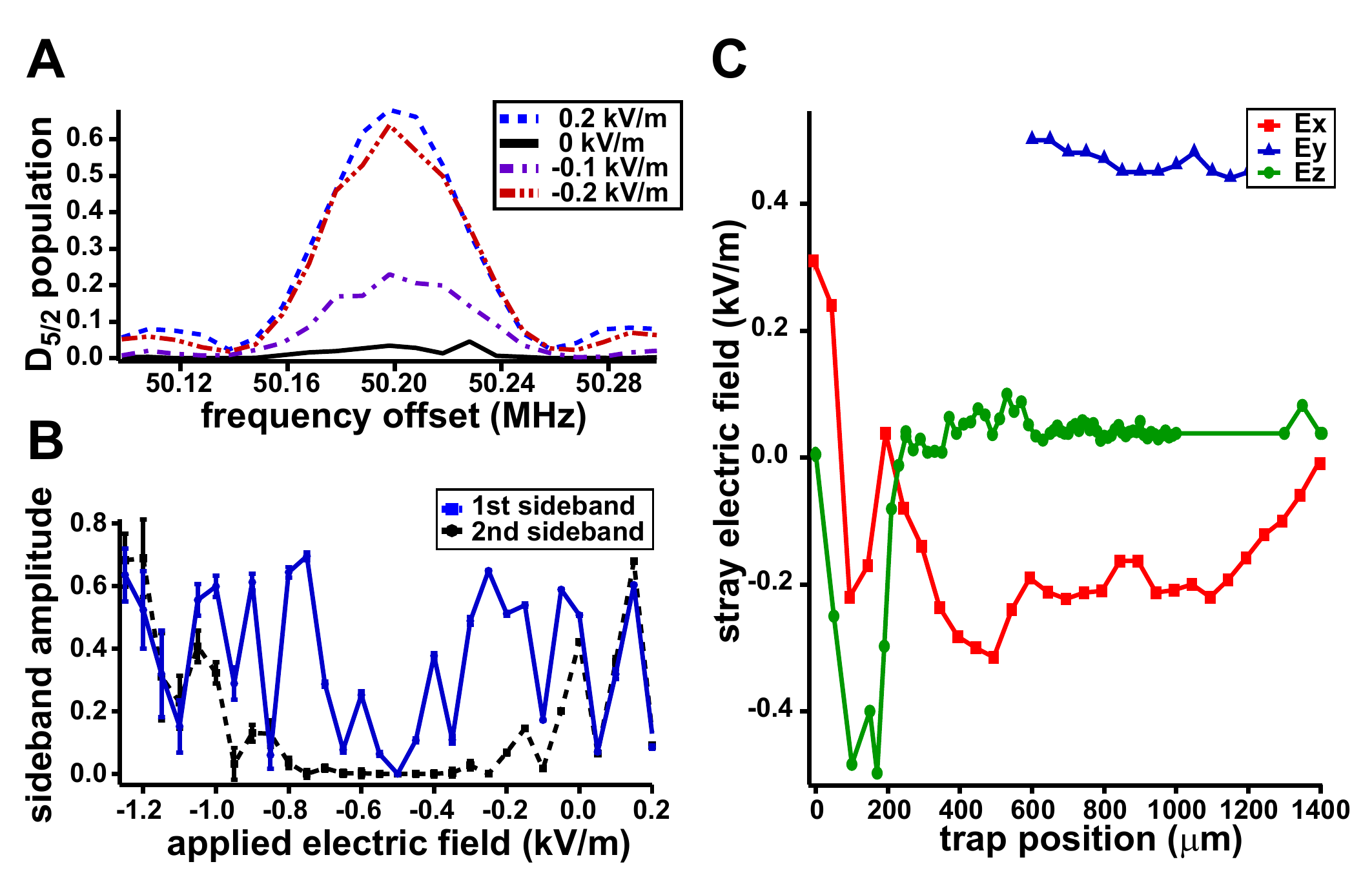}
\caption{Compensation of stray electric fields.  (a) Micromotion sidebands may be probed perpendicular to the trap plane using a laser tuned to the $S_{1/2} \rightarrow D_{5/2}$ transition at 729~nm.  (b) Stray electric fields are compensated when both the first- and second-order sidebands are simultaneously minimized.  (c) Measured stray electric fields in three dimensions.}
\label{fig:y-axis_comp}
\end{figure}

Stray electric fields along $\hat{z}$ are easily measured by monitoring the ion position while rescaling the axial harmonic well strength, and micromotion along $\hat{x}$ may be minimized by observing resolved micromotion sidebands.\cite{Doret12, Peik99}  However, in surface-electrode traps it is more challenging to minimize micromotion along $\hat{y}$.  This requires that there be overlap between the probe laser's propagation direction and the $y$-axis.  This leads to unacceptable levels of light scatter and charging of dielectrics by the blue fluorescence lasers generally used.  Two principle alternatives have been developed for measuring $\hat{y}$ micromotion, including orienting the infrared repumping lasers along this direction\cite{Allcock10} and parametrically exciting the ion.\cite{Ibaraki11, Narayanan11}  Here we use a third alternative: monitoring micromotion sidebands on the $S_{1/2} \rightarrow D_{5/2}$ electric quadrupole transition at 729~nm, for which surface charging effects are negligible.\cite{Harlander10, Wang11}  A 729~nm laser is aligned onto the ion by projecting it backwards down the fluorescence collection optics via a dichroic mirror (Semrock FF670-SDi01-25x36).  As this aligns the beam along $\hat{y}$, micromotion in this direction may be easily minimized by monitoring first- and second-order micromotion sidebands and adjusting applied compensation fields~(Fig.~\ref{fig:y-axis_comp}).  

\subsection{Ion Heating}
\begin{figure}
\includegraphics[width=8.0 cm]{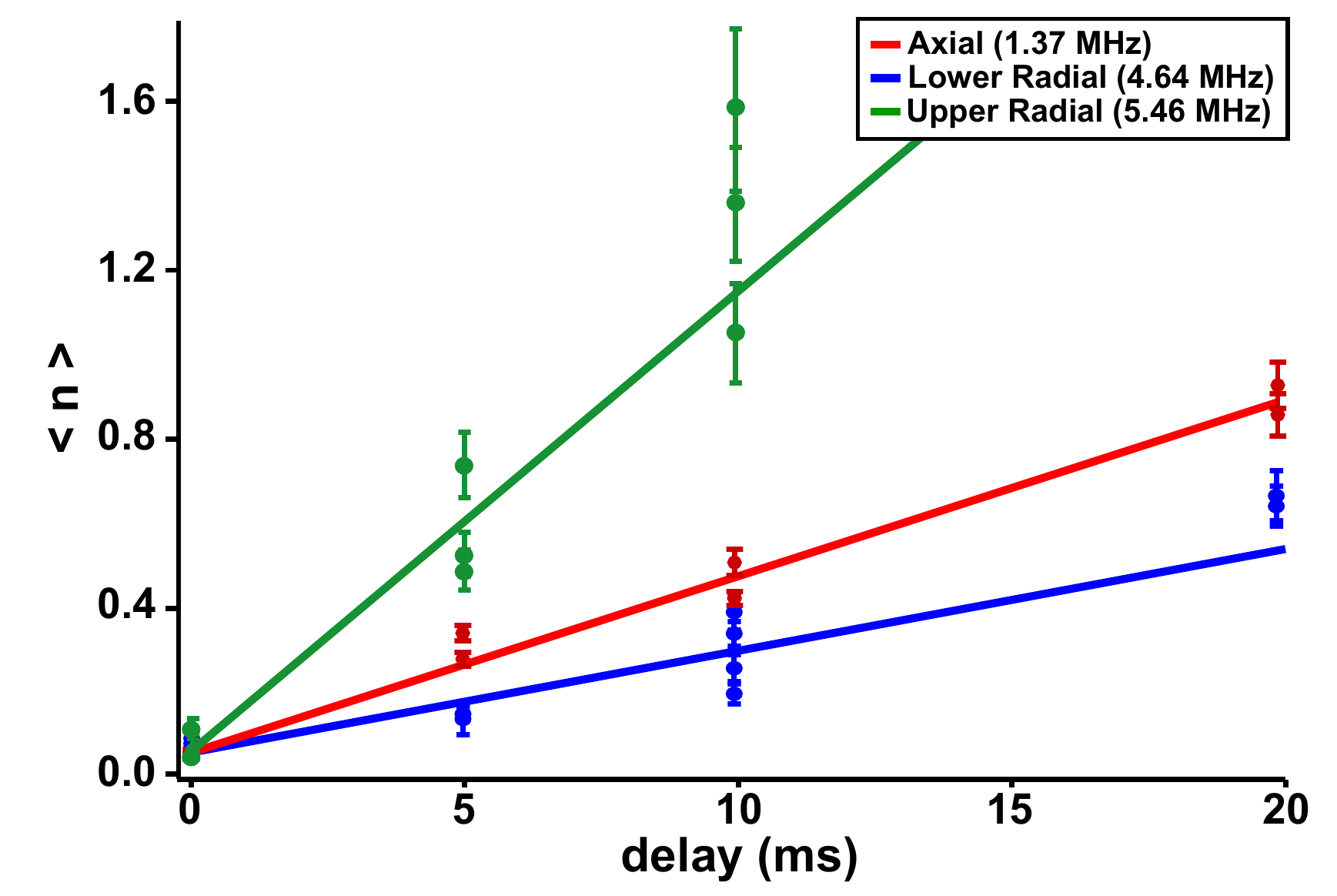}
\caption{Ion heating.  Fitted heating rates are 41(3), 24(3), and 110(3) quanta/sec for the axial, lower radial, and upper radial secular modes, respectively.}
\label{fig:heating}
\end{figure}

Along with improved vacuum, reduced ion heating rates are the dominant motivation for cryogenic ion trapping systems.  We measure the heating of a single trapped ion from the ground state for all three secular modes $\omega_{z} = 1.38$~MHz, $\omega_{r_{1}} = 4.64$~MHz, $\omega_{r_{2}} = 5.46$~MHz, finding rates of 41(3), 24(3), and 110(3) quanta/sec, respectively (Fig.~\ref{fig:heating}).  As similar measurements\cite{Turchette00} elsewhere indicate that heating rates scale as $1/\omega^{2}$, it seems likely that the RF trap drive is inadequately filtered by the helical resonator.  This would predominantly increase heating of the radial modes, thus we take the axial heating rate as an upper bound on the trap's spectral noise density, $S(E) < 3.8(3) \times 10^{-13}$~V$^{2}/$m$^{2}$ Hz, a reduction of approximately an order of magnitude over an identical trap measured at room temperature.\cite{Doret12} 

\section{Conclusion}
We have developed a modular cryostat for use with surface electrode traps.  A commercial cryocooler with a mechanically isolated auxilliary mounting stage provides large cooling power, low base temperatures, and motion of less than 5 (1)~$\mu$m at full (35$\%$) duty cycle.   The system uses an O-ring sealed vacuum enclosure with a single vacuum space and only one thermal shield, and does not require special accomodations to heat sink the ion trap.   As such, the system is amenable to the use of standard room temperature ion trap technologies, including resistive oven sources and UHV compatible trap packaging, allowing traps to be used interchangably between low-temperature and room temperature systems.  Despite these simplifications over previous cryogenic ion trapping systems, ion lifetimes and heating rates are both improved by more than an order of magnitude as compared to an identical trap at room temperature.

Simple cryogenic ion trapping systems such as this are well suited to future experiments requiring low ion heating rates or exceptional vacuum.  For example, suppression of motional heating, a leading cause of decoherence for motional coupling of ions in adjacent potential wells,\cite{Brown11, Harlander11} could assist in the generation of entangled pairs of ions for use in a large quantum computer.\cite{Kielpinski02}  Similarly, avoiding ion-chain reordering induced by background gas collisions is essential for schemes involving co-trapping of cooling and computational ions.\cite{Lin09, Lin11}  In addition, the low temperature environment helps to bridge the gap between different quantum technologies such as trapped ion and superconducting qubits, and may further facilitate the construction of hybrid quantum systems by enabling the use of materials otherwise incompatible with UHV conditions.



%
%

%

\begin{acknowledgments}
We would like to thank David Patterson and Amar Vutha for helpful design suggestions and Chris Shappert for assistance with thermal modeling.  This work has been funded by the Georgia Tech Research Institute.  GDV and SCD thank GTRI Shackelford Fellowships for graduate and postdoctoral support, respectively.  
\end{acknowledgments}

\bibliographystyle{aipnum4-1}

\bibliography{cryoRSIbib}

\begin{thebibliography}{42}
\bibitem [{hea()}]{heatsink}%
  \BibitemOpen
  \href@noop {} {}\bibinfo {note} {The trap and CPGA carrier are thermally
  anchored via the DC electrical connections. At the time of the trap
  temperature measurement, the DC feeds were heatsunk via the DC wiring only,
  approximately 15~cm distant. We have since added a copper heatsink to the
  fan-out PCB, thermally anchoring the DC traces adjacent to the trap mounting
  socket. This should reduce the thermal gradient between the CPGA and
  cold-plate, dropping the trap temperature to $\sim$ 8~K under normal
  operating conditions.}\BibitemShut {Stop}%
\end{thebibliography}

\end{document}